\documentclass[prb,preprint]{revtex4}

\usepackage{amsmath}    
\usepackage{amsfonts}  
\usepackage{amssymb}
\usepackage{graphicx}   

\usepackage{dcolumn}
\usepackage{bm}

\newcommand{\be}{\begin{equation}}
\newcommand{\ee}{\end{equation}}
\newcommand{\bea}{\begin{eqnarray}}
\newcommand{\eea}{\end{eqnarray}}

\newcommand{\rme}{{\rm{e}}}

\newcommand{\iGA}{{j}}
\newcommand{\iGAT}{{i}}

\newcommand{\ex}{{e_1}}
\newcommand{\ey}{{e_2}}
\newcommand{\ez}{{e_3}}

\newcommand{\reversion}[1] { {#1}^{\dagger} }
\newcommand{\cliffconj}[1] { \bar{#1} }

\begin{document}


\title{Geometric Algebra: A natural representation of three-space}  
\author{James M.~Chappell}
\email{james.chappell@adelaide.edu.au}
\affiliation{School of Electrical and Electronic Engineering, University of Adelaide, SA
5005, Australia}
\author{Azhar Iqbal}
\affiliation{School of Electrical and Electronic Engineering, University of Adelaide, SA
5005, Australia}
\author{Derek Abbott}
\affiliation{School of Electrical and Electronic Engineering, University of Adelaide, SA
5005, Australia}
\date{\today}

\begin{abstract}
Historically, there have been many attempts to produce an appropriate mathematical formalism for modeling the nature of physical space, such as Euclid's geometry, Descartes' system of Cartesian coordinates, the Argand plane, Hamilton's quaternions and Gibbs' vector system using the dot and cross products.  We illustrate however, that Clifford's geometric algebra (GA) provides the most elegant description of physical space.  Supporting this conclusion, we firstly show how geometric algebra subsumes the key elements of the competing formalisms and secondly how it provides an intuitive representation of the basic concepts of points, lines, areas and volumes. We also provide two examples where GA has been found to provide an improved description of two key physical phenomena, electromagnetism and quantum theory, without using tensors or complex vector spaces. This paper also provides pedagogical tutorial-style coverage of the various basic applications of geometric algebra in physics.
\end{abstract}

\maketitle

\section{Introduction}

Einstein once stated, {\it{`Everything should be made as simple as possible, but not one bit simpler'}}, and in this paper we ask the question: `What is the simplest mathematical representation of three-dimensional physical space that is nevertheless complex enough to satisfactorily describe all its key properties?'

The presence of exactly five regular solids leads to the conclusion that we live in a three-dimensional world. If we lived in a world with four spatial dimensions, for example, we would be able to construct six regular solids, and in five dimensions and above we would find only three \cite{coxeter1973regular}. Also, the gravity and the electromagnetic force laws have been experimentally verified to follow an inverse square law to very high precision \cite{hoyle2001submillimeter}, indicating the absence of additional macroscopic dimensions beyond three space dimensions.
Hence a three-dimensional coordinate system, as proposed by Descartes, appears to be a good starting point to describe physical space. For three-space, however, as well as positional coordinates, we realize that we also need to be able to represent rotations at each point in this space. In the plane, the algebra of complex numbers can be used for rotations and in three space, the algebra for rotations is given by Hamilton's quaternions.  
Hence, in order to form a unified algebra of three-space we need to integrate the algebra of the complex numbers and quaternions within the framework of Cartesian coordinates. This was achieved by Clifford in 1873, who named his system, {\it Geometric Algebra} (GA).

\subsection{Historical development}

Pythagoras famously stated that numbers and their relationships, underlie all things.  This mathematical idealism of Pythagoras was masterfully applied by Euclid, to geometry, deriving his famous set of geometrical theorems based on a few simple axioms that formed the first comprehensive theory for the physical world  \cite{Wedberg1982}.  
The next major breakthrough in mathematical science did not come though till the seventeenth century and it has been extensively debated by historians, why there was such a slow down in the progress of science and mathematics following the Greek explosion.  Various suggestions have been provided to answer this, such as the Roman empire suppressing dissent and not sponsoring the arts \cite{Russel1945}, but more recently it has been proposed, that the algebraic and numerical system used by the Greeks, had inherent limitations, which were roadblocks to further progress \cite{Hestenes111}  and illustrates in this case the importance of the mathematical framework utilized in order to promote scientific progress.
For example, in the Greek system, the length along the diagonal of a unit square, we know today as $ \sqrt{2} $, being an irrational number, did not exist in the Greek numeric system, which was based solely on integers and their ratios. 

However with the arrival of Hindu-Arabic numbers in about 1000~AD into Europe, which included a zero that allowed positional representation for numbers, together with the acceptance of negative numbers in 1545~AD, allowed for the concept of a complete number line to be developed.  This then paved the way for Descartes to revolutionize the Greek system in 1637, by proposing a union of algebra and geometry using Cartesian coordinates.  He stated {\it{`Just as arithmetic consists of only four or five operations, namely, addition, subtraction, multiplication, division and the extraction of roots, which may be considered a kind of division, so in geometry, to find required lines it is merely necessary to add or subtract lines.'}}
Descartes thus postulated an equivalence between line segments and numbers, something the Greeks were not prepared to do.  
This achievement is identified by  John Stuart Mill, {\it{`the greatest single step ever made in the exact sciences'}} \cite{StuartMill1979}.

The Cartesian coordinate system proposed by Descartes, becomes confused, however, with the later development of the Argand diagram, which, while isomorphic to the Cartesian plane, consists of one real and one imaginary axis, and so not compatible with the idea of space as isotropic. To add to the confusion, Hamilton in 1843 generalized the complex numbers to three space, defining the algebra of the quaternions using the basis elements $ \rm{i}, \rm{j},\rm{k} $ that can also be used as a basis for three dimensional space.  
This confused state of affairs, on exactly how to represent three-space coordinates and rotations, was finally resolved by William Clifford in 1873.  Clifford accepted the Cartesian coordinate system of Descartes, but then integrated the algebra of complex numbers and quaternions as the rotation operators within this space.  
Clifford also achieved a fulfillment of Descartes' original vision of a vector being able to be manipulated
in the same way as ordinary algebraic quantities, by deriving a multiplication and division operation for vectors.  Additionally Clifford's system extended this idea to its natural conclusion, allowing not just lines, but also areas and volumes to be also treated in this same way. 
Clifford however also extended this idea by claiming that {\it{`That this variation of the curvature of space is what really happens in that phenomenon which we call the motion of matter, whether ponderable or ethereal.'}} \cite{Clifford1876} This foreshadowed the idea used by Einstein many years later for his theory of gravity. The properties of spacetime as espoused in general relativity were nicely expressed by Drude \cite{drude1920}: {\it{`The conception of ether absolutely at rest is the most simple and the most natural, at least if the ether is conceived to be not a substance but merely space endowed with certain physical properties.'}}   

\subsection{Clifford's definition of three-space}

\begin{figure}[htb]
\begin{center}
\includegraphics[width=3.5in]{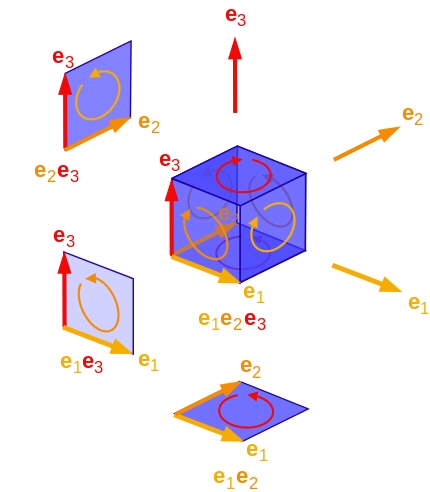}
\end{center}
\caption{The elements of Clifford's model for three space. This consists of three unit vectors $ e_1, e_2 $ and $ e_3 $, three unit areas $ e_2 e_3, e_3 e_1 $ and $ e_1 e_2 $ and a unit volume $ \iGA = e_1 e_2 e_3 $. The pure scalars then defining points to form a complete algebraic description of three-dimensional physical space. \label{ThreeSpace}}
\end{figure}

How did Clifford solve the problem of forming an integrated description of three-space that combined Cartesian coordinates and the algebra of complex numbers and quaternions? 
Following Clifford, we firstly represent the three degrees of freedom in a Cartesian coordinate system by the algebraic constants $ e_1, e_2 $ and $ e_3 $ as shown in Fig.~\ref{ThreeSpace}, which we define to have a positive square, that is $ e_1^2 = e_2^2 = e_3^2 = 1$. The next crucial step is then to specify these elements as anticommuting, that is $ e_j e_k = - e_k e_j $ for $ j \ne k $.  These few definitions are sufficient to define Clifford's system. 

Geometrically, the basis elements $ e_1, e_2,e_3 $, the bivectors $ e_2 e_3 $, $ e_1 e_3 $ and $ e_2 e_3 $ and the trivector $ e_1 e_2 e_3 $, are natural constructs to represent unit lines, unit areas, and unit volumes respectively.
We also find that the compound algebraic elements, the bivectors $ e_2 e_3 $, $ e_1 e_3 $ and $ e_2 e_3 $ all square to minus one, for example, $ (e_1 e_2)^2 = e_1 e_2 e_1 e_2 = -e_1 e_1 e_2 e_2 = -1 $, using the anticommutivity and positive square of the basis elements.  We can now identify an isomorphism of the three bivectors with the three quaternions of Hamilton, so that $ {\rm{i}} \leftrightarrow e_2 e_3 $, $ {\rm{j}} \leftrightarrow e_1 e_3 $, $ {\rm{k}} \leftrightarrow e_1 e_2 $,thus integrating the quaternions into Clifford's system. Also, in the plane, the bivector $ e_1 e_2 $ can be used as a replacement for the unit imaginary $ \sqrt{-1} $, forming a complex-like number $ a + i b $, where we define $ \iGAT = e_1 e_2 $.
The final compound element, the trivector $ \iGA = e_1 e_2 e_3 $ also squares to minus one and commutes with all basis elements and so is isomorphic to the scalar unit imaginary $ \sqrt{-1} $ in three dimensions.  Now, because the unit imaginary is no longer required in Clifford's system, and because the unit imaginary was first used in complex numbers that are isomorphic to GA in two dimensions, we will adopt the widely used symbol $ \iGAT = e_1 e_2 $ to represent the unit imaginary when in two dimensions, and in three dimensions, we will adopt $ \iGA = e_1 e_2 e_3 $, a commonly used symbol in electrical engineering to represent the unit imaginary. This distinction between two forms of the unit imaginary $ \sqrt{-1} $, as $ \iGAT $ and $ \iGA $ in two and three dimensions respectively, has physical significance that we illustrate when describing Dirac's equation for the electron, in a later section.

Now, using the trivector $ \iGA $ we also find the dual relations $ e_1 e_2 = \iGA e_3 $, $ e_3 e_1 = \iGA e_2 $ and $ e_2 e_3 = \iGA e_1 $.  For example, $ \iGA e_1 = e_1 e_2 e_3 e_1 = e_1^2 e_2 e_3 = e_2 e_3 $, as required. These relations can be summarized by the relation $ e_p e_q = \delta_{pq} + \iGA \epsilon_{pqr} e_r $, where $ p,q,r \in \{1,2,3\} $ and $ \delta , \epsilon $ are the well known  Kronecker delta function's and the antisymmetric tensor respectively, which we see describes the Pauli algebra, in fact. Hence, we can use Clifford's basis vectors $ e_k $ to replace the three Pauli matrices $ \sigma_k $ commonly used to describe quantum mechanical spin. 

\begin{figure}[htb]

\begin{center}
\includegraphics[width=4.8in]{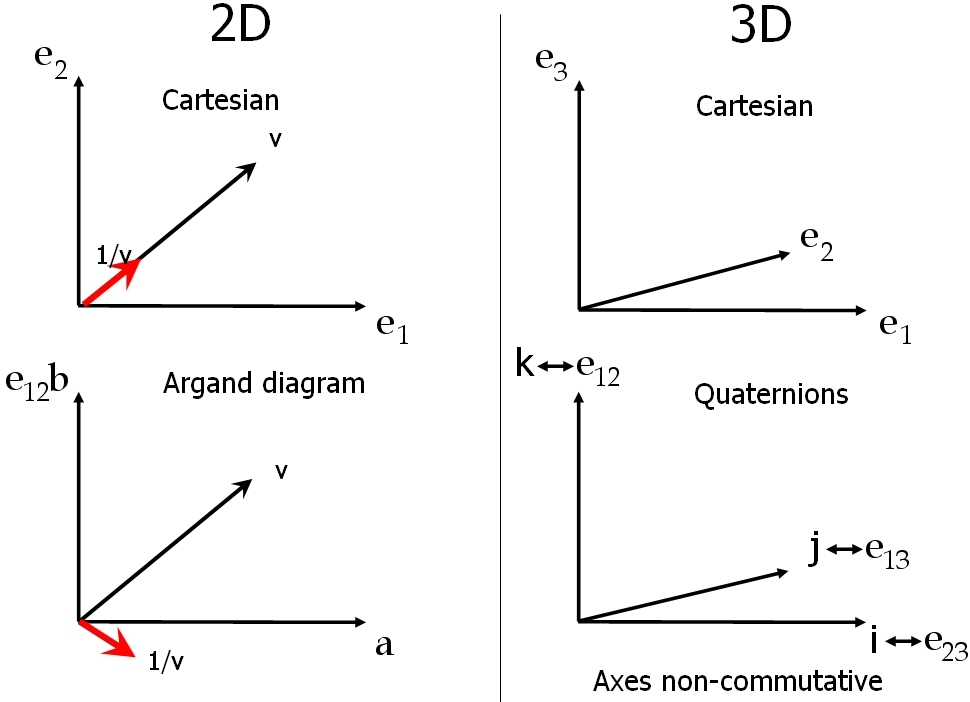}
\end{center}

\caption{Clifford's representation of three space. Firstly, the Cartesian plane is defined by the vectors $ e_1 $ and $ e_2 $, with the bivector $ e_{12} $ identified with the unit imaginary used to define the Argand diagram. The inverse of a vector $ \boldsymbol{v} $ is also shown in red, both for the Cartesian case and for the Argand plane. In 3D, defined by the vectors $ e_1, e_2 $ and $ e_3 $, the non-commuting quaternions $ \rm{i},\rm{j} $ and $ \rm{k} $ are replaced by the three bivectors $ e_{23} $, $ e_{13} $ and $ e_{12} $ as shown.  \label{CoordSystems}}
\end{figure}
The competing mathematical systems that Clifford unified are shown in Fig.~\ref{CoordSystems}.  In the plane the unit imaginary $ \sqrt{-1} $ is replaced with the bivector $ \iGAT = e_{12} $, using the subscript notation $ e_{12} =  e_1 e_2 $.  
Hamilton's three quaternions $ \rm{i},\rm{j} $ and $ \rm{k} $ representing rotations about the three available axes, can be replaced with the bivectors $ e_{23} $, $ e_{13} $ and $ e_{12} $ respectively, as shown, with the Cartesian axes described by unit vectors $ e_1 $, $ e_2 $ and $ e_3 $. Thus Clifford's system, consisting of just the three elements $ e_1 , e_2 $ and $ e_3 $, unifies Cartesian coordinates,  complex numbers and the quaternions into a single mathematical system.

\subsection{The Clifford vector product}

Using the three basis elements we can define a vector $ \boldsymbol{v} = v_1 e_1 + v_2 e_2 + v_3 e_3 $, where $ v_i \in \Re $, and given a second vector $ \boldsymbol{u} = u_1 e_1 + u_2 e_2 + u_3 e_3 $, we can find their algebraic product using the distributive law of multiplication over addition, giving
\bea \label{VectorProductExpand}
& & \boldsymbol{u} \boldsymbol{v} \\ \nonumber
& = & (e_1 u_1 + e_2 u_2 + e_3 u_3 ) ( e_1 v_1 + e_2 v_2 + e_3 v_3 ) \\ \nonumber
& = & u_1 v_1 + u_2 v_2 + u_3 v_3 + (u_2 v_3 - v_2 u_3 ) e_2 e_3 + (u_1 v_3 - u_3 v_1 ) e_1 e_3 + (u_1 v_2 - v_1 u_2 ) e_1 e_2 \\ \nonumber
& = & \boldsymbol{u} \cdot \boldsymbol{v}  + \boldsymbol{u} \wedge \boldsymbol{v}, \nonumber
\eea
which produces a sum of the dot and wedge products.  This algebraic product is commonly referred to as the geometric product.  If we now use the dual relations established earlier,  we can transform this result to
\bea \label{CrossProductDefn}
\boldsymbol{u} \boldsymbol{v} & = & u_1 v_1 + u_2 v_2 + u_3 v_3 + \iGA \left ((u_2 v_3 - v_2 u_3 ) e_1 + (u_1 v_3 - u_3 v_1 ) e_2 + (u_1 v_2 - v_1 u_2 ) e_3 \right ) \\ \nonumber
& = & \boldsymbol{u} \cdot \boldsymbol{v}  + \iGA \boldsymbol{u} \times \boldsymbol{v}, \nonumber
\eea
which now forms a resultant in the form of a complex-like number consisting of the dot and cross products.   Both the wedge product form and the cross product form are useful, though the dual relation, allowing the cross product form, only applies in  three dimensions.  Hence we can see that the dot and the cross products indeed appear intrinsic to three dimensional space, however the advantage of the Clifford system is that they are unified into a single invertible number.  We can also identify a limitation of defining the cross product as a separately defined product, as it does not naturally extend to higher dimensional spaces, whereas the formulation in Eq.~(\ref{VectorProductExpand}) does.  The expression in Eq.~(\ref{CrossProductDefn}) generated by simply expanding the brackets defining two vectors thus provides an alternative calculation tool to the conventional method of calculating the determinant of a $ 3 \times 3 $ matrix formed from the components of the two vectors.  Clifford's description of vector products relying solely on the elementary rules of algebra, thus has the advantage of avoiding the additional mathematical machinery required when handling traditional row or column vectors.

As can be seen from Eq.~(\ref{VectorProductExpand}), for the case of a vector multiplied by itself, the wedge product will be zero and hence the square of a vector $ \boldsymbol{v}^2 = \boldsymbol{v} \cdot \boldsymbol{v}  $, becomes a scalar quantity.
This then allows us to define the inverse of a vector $ \boldsymbol{v} $ as 
\be \label{vectorInverse}
\boldsymbol{v}^{-1} = \frac{1}{\boldsymbol{v}^2 } \boldsymbol{v} .
\ee
Checking this result we find $ \boldsymbol{v} \boldsymbol{v}^{-1} = \frac{1}{\boldsymbol{v}^2 } \boldsymbol{v} \boldsymbol{v} = 1 $ as required, so that we can form the vector division $ \frac{\boldsymbol{u}}{\boldsymbol{v}} \equiv \boldsymbol{u} \boldsymbol{v}^{-1} $.  

We can now compare the inverse of a Cartesian vector with the inverse of complex number. Given a complex number $ z = r \rme^{\iGAT \theta } $ we find the inverse $ z^{-1} = (1/r) \rme^{- \iGAT \theta  } $, that has an inverse length, with a negative angle. For a Cartesian vector $ \boldsymbol{v} = e_1 r \rme^{ \iGAT \theta } = r \cos \theta e_1 + r \sin \theta e_2 $ in Clifford's system, we find the inverse vector $ \boldsymbol{v}^{-1} = (1/r)  e_1 \rme^{ \iGAT \theta } =  (1/r) \cos \theta e_1 + (1/r) \sin \theta e_2 $ that is a vector of inverse length, but in the same direction as the original vector, as shown in Fig.~\ref{CoordSystems}. The negative direction for the angle $ \theta $ for the case of the inverse of a complex number, forming an inverse rotation, also confirms their natural role as rotation operators rather than as a replacement for Cartesian vectors. 

\begin{figure}[htb]
\begin{center}
\includegraphics[width=2.8in]{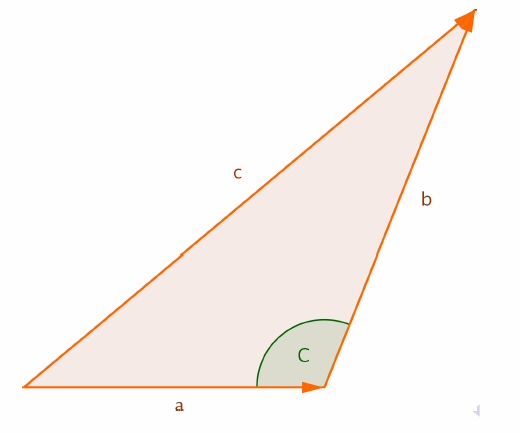}
\end{center}
\caption{Deriving the law of cosines for triangles. From the diagram we have the vector relation that $ \boldsymbol{c} = \boldsymbol{a} + \boldsymbol{b} $, that gives using the geometric product $ \boldsymbol{c}^2 = \boldsymbol{a}^2 +  \boldsymbol{b}^2 + 2 \boldsymbol{a} \cdot \boldsymbol{b} $.  \label{CosRule}}
\end{figure}

As a simple application of Clifford's geometric product, given two vectors $ \boldsymbol{a} = a_1 e_1 + a_2 e_2 + a_3 e_3 $ and $ \boldsymbol{b} = b_1 e_1 + b_2 e_2 + b_3 e_3 $, we can produce a third vector $ \boldsymbol{c} = \boldsymbol{a} + \boldsymbol{b} $, as shown in Fig.~\ref{CosRule}.
We then find that
\be
\boldsymbol{c}^2 = ( \boldsymbol{a} + \boldsymbol{b})^2 = \boldsymbol{a}^2 +  \boldsymbol{b}^2 + \boldsymbol{a} \boldsymbol{b} + \boldsymbol{b} \boldsymbol{a} =  \boldsymbol{a}^2 +  \boldsymbol{b}^2 + 2 \boldsymbol{a} \cdot \boldsymbol{b}  ,
\ee
using the the result from Eq.~(\ref{CrossProductDefn}) that $ \boldsymbol{a} \boldsymbol{b} + \boldsymbol{b} \boldsymbol{a} = \boldsymbol{a} \cdot \boldsymbol{b} + j \boldsymbol{a} \times \boldsymbol{b} + \boldsymbol{b} \cdot \boldsymbol{a} + j \boldsymbol{b} \times \boldsymbol{a} = 2 \boldsymbol{a} \cdot \boldsymbol{b} = - 2 |\boldsymbol{a}| |\boldsymbol{b}  \cos C $, where $ C $ is the angle between the vectors $ \boldsymbol{a} $ and $ \boldsymbol{b} $, a result also known as the law of cosines for triangles.  Note the minus sign in the relation $ \boldsymbol{a} \cdot \boldsymbol{b} = - |\boldsymbol{a}| |\boldsymbol{b}  \cos C $ as the angle defined by $ \boldsymbol{a} \cdot \boldsymbol{b} $ produces the exterior angle $ \pi - C $ in this case.

\subsection{Example 1: Area calculation}

\begin{figure}[htb]

\begin{center}
\includegraphics[width=4.8in]{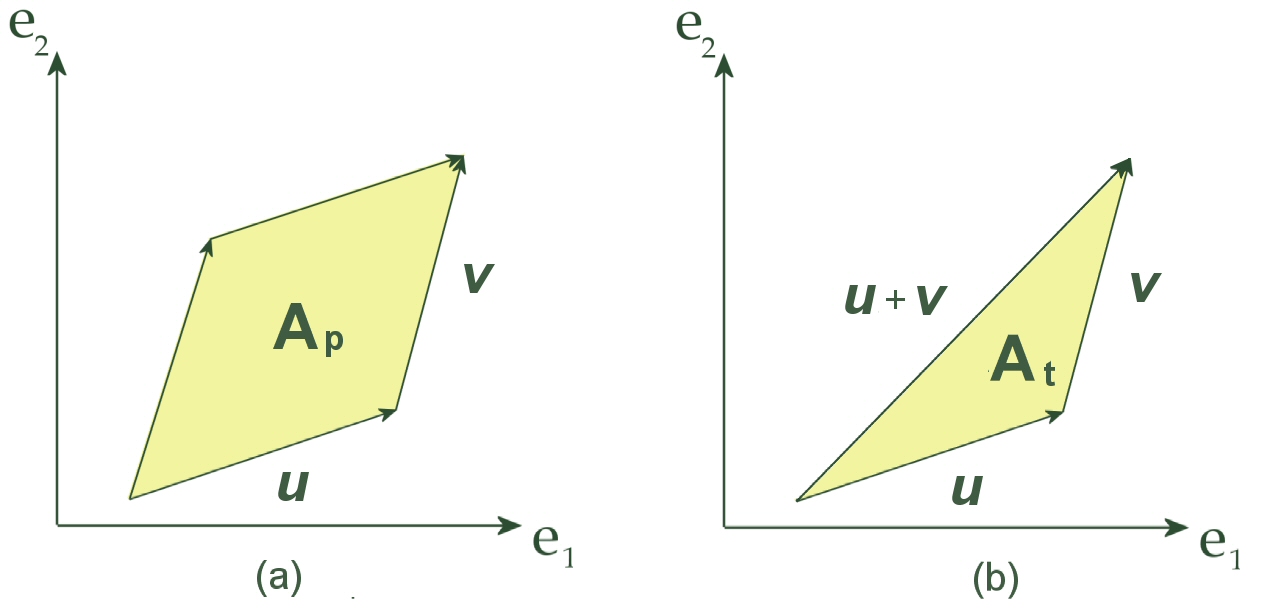}
\end{center}

\caption{Calculating areas using the geometric product. For two vector $ \boldsymbol{u} $ and $ \boldsymbol{v} $ we can find the area of the parallelogram $ A_p = \langle \boldsymbol{u} \boldsymbol{v} \rangle_2 $ and the area of the triangle $ A_t = \frac{1}{2} \langle \boldsymbol{u} \boldsymbol{v} \rangle_2 $, as shown. \label{TwoVectors}}

\end{figure}

Inspecting Fig.~\ref{TwoVectors}(a), we might wish to know the area enclosed by the two vectors, which we can calculate from a variety of geometrical constructions, to be $ u_1 v_2 - u_2 v_1 $.  Alternatively, from Eq.~(\ref{VectorProductExpand}), we can write the product of the two vectors
\be \label{geoProductFull}
\boldsymbol{u} \boldsymbol{v} = \boldsymbol{u} \cdot \boldsymbol{v} + \boldsymbol{u} \wedge \boldsymbol{v} = u_1 v_1 + u_2 v_2 + (u_1 v_2 - u_2 v_1) e_1 e_2 ,
\ee
then we can see that the area is given by the bivector term $ \boldsymbol{u} \wedge \boldsymbol{v} $.  The bivector $ e_1 e_2 $ represents a unit area, and so it is natural to expect this component to represent the area.  Therefore we can write for the area of the parallelogram
\be
A_p = \langle \boldsymbol{u} \boldsymbol{v} \rangle_2 ,
\ee
where the notation $ \langle \boldsymbol{u} \boldsymbol{v} \rangle_2 $ means to retain the second grade or bivector terms.  Dimensionally this also makes sense, because we are looking for a result with dimensions of area or squared length.
This argument also applies to three dimensions, where the volume will therefore need to be grade 3, that is for three vectors we find the enclosed volume $ V = \langle \boldsymbol{u} \boldsymbol{v} \boldsymbol{w} \rangle_3 $ as expected.
Thus a routine calculation of the geometric product, followed by the selection of the desired components dimensionally, allows the relevant information to be extracted.

Consequently, if we wish to find the area of a triangle with two sides given by $ \boldsymbol{u} $ and $ \boldsymbol{v} $ where the third side is then given by $ \boldsymbol{u} + \boldsymbol{v} $, as shown in Fig.~\ref{TwoVectors}(b), then clearly we form an area $ A_t = \frac{1}{2} \langle \boldsymbol{u} \boldsymbol{v} \rangle_2 $.  Picking any two sides of the triangle will produce the same area, that is $ \boldsymbol{u} \wedge \boldsymbol{v} = (\boldsymbol{u} + \boldsymbol{v}) \wedge  \boldsymbol{v} = \boldsymbol{u} \wedge  (\boldsymbol{u} + \boldsymbol{v} ) $ as $ \boldsymbol{u} \wedge \boldsymbol{u} = \boldsymbol{v} \wedge \boldsymbol{v} =  0 $.
This approach to calculating areas can also be extended to more general shapes simply by dividing the shape into a set of triangles and then summing the triangles.

\subsection{The multivector}

In GA, the basis elements $ e_1, e_2 $ and $ e_3 $ are algebraic constants and so we are free to add the various scalar, vector, bivector and trivector components together.
In fact, adding all available components, we form the space of multivectors $ \Re \oplus \Re^3 \oplus \bigwedge^2 \Re^3 \oplus \bigwedge^3 \Re^3 $, an eight-dimensional real vector space also denoted by $ C\ell_{3}(\Re) $, which can be written
\be \label{generalMultivector}
M = a + \boldsymbol{v} + \iGA \boldsymbol{w} + \iGA t ,
\ee
which shows in sequence, a scalar $ a $, vector $ \boldsymbol{v} = v_k e_k $, bivector $ \iGA \boldsymbol{w} = \iGA w_k e_k $ and trivector $ \iGA t $ terms, where $ k \in \{1,2,3\} $.
This general three-space multivector can be used to represent lines, areas and volumes within three dimensions, as well as a diverse range of physical phenomena such as electromagnetic fields.
As already noted the bivectors are isomorphic to the three quaternions $ \rm{i},\rm{j},\rm{k} $, and we find the multivector $ a + \iGA \boldsymbol{b} = a + \iGA b_1 e_1 + \iGA b_2 e_2 + \iGA b_3 e_3 $ isomorphic to a quaternionic number $ q = a + b_1 \rm{i} - b_2 \rm{j} + b_3 \rm{k} $. 
We have already identified $ e_1, e_2 $ and $ e_3 $ with the Pauli matrices, and for a Pauli spinor, representing a spin-$\frac{1}{2} $ particle, we have the mapping 
\be \label{SpinorMapping}
| \psi \rangle = \begin{bmatrix} a + {\rm{i}} a_3  \\ -a_2 + {\rm{i}} a_1  \end{bmatrix} \leftrightarrow \psi = a + \iGA a_1 e_1 + \iGA a_2 e_2 + \iGA a_3 e_3 = a + \iGA \boldsymbol{a},
\ee
where $ {\rm{i}} = \sqrt{-1} $, also mapping to the even sub algebra of the multivector, which shows the equivalence of GA bivectors, Pauli spinors and quaternions \cite{CIL}.

We will also see later how the electromagnetic field antisymmetric tensor $ F^{\mu \nu} $, maps as follows into the vector and bivector components of the multivector \cite{Griffiths:1999}
\be \label{FieldTensorMapping}
F^{\mu \nu} \leftrightarrow F = \boldsymbol{E}+\iGA \boldsymbol{B} ,
\ee
with the dual tensor $ G^{\mu \nu} $ given in GA by simply $ G = \iGA F $. We have found that the scalars and bivectors can be used to represent the Pauli spinors, and the vector and bivector components used to describe the electromagnetic field and so we might ask if there is any physical phenomena that requires the full multivector, as shown in Eq.~(\ref{generalMultivector}) for its representation.  We find, in fact, that the wave function used to represent the electron in Dirac's relativistic wave equation, maps to the full multivector.

\begin{table}
	\centering

\begin{tabular}{|l|l|l|l|}
\hline
 GA multivector & Alternate formalism & Description   \\
\hline \hline
$ v_1 e_{1} + v_2 e_{2} + v_3 e_{3} $ & $ [ \sigma_1, \sigma_2, \sigma_3 ]^T $ &  Vectors/Pauli matrices  \\
$ a + \iGAT b $, $ \iGAT = e_{12} $ & $ a + {\rm{i}}  b $, $ \,\,\, {\rm{i}} = \sqrt{-1}$ & Complex numbers     \\
$ a + b e_{23} + c e_{13} + d e_{12} $ & $ a + b {\rm{i}} + c {\rm{j}} + d {\rm{k}}  $, $ | \psi \rangle = \begin{bmatrix} a + {\rm{i}} a_3  \\ -a_2 + {\rm{i}} a_1  \end{bmatrix} $ &  Quaternions/Pauli spinors  \\
$ \boldsymbol{E} + j \boldsymbol{B} $ & $ F^{\mu \nu },\, \mu,\nu \in \{0,1,2,3\} $  &  Electromagnetic field tensor     \\
$ a + \boldsymbol{E} + j \boldsymbol{B} + j b $ & $ \psi_{\mu}, \,\,\, \mu \in \{0,1,2,3\} $  &  Complex Dirac wave function   \\
 
\hline
\end{tabular}
\caption{An illustration of how the GA multivector replaces a variety of alternate mathematical formalisms. }    \label{tableFormalisms}
\end{table}

The great versatility of the three-space multivector is demonstrated in Table \ref{tableFormalisms}, being able to replace a large variety of mathematical structures and formalisms, as well as elegantly describe many physical phenomena~\cite{GA2}.

\subsection{Common algebraic operations on a multivector}

Descartes claimed that the five common algebraic operations of addition, subtraction, multiplication, division and square root, could be applied to his line segments, however this idea can now be extended to areas and volumes as well as composite quantities described by the multivector in Eq.~(\ref{generalMultivector}).  The multivector represents a set of elements, containing a line element, an areal element (bivector) and a volume element (trivector).  When algebraic operations are applied to these sets of geometric elements, we form a new set of geometric elements within the space of multivectors.

\begin{figure}[htb]

\begin{center}
\includegraphics[width=3.4in]{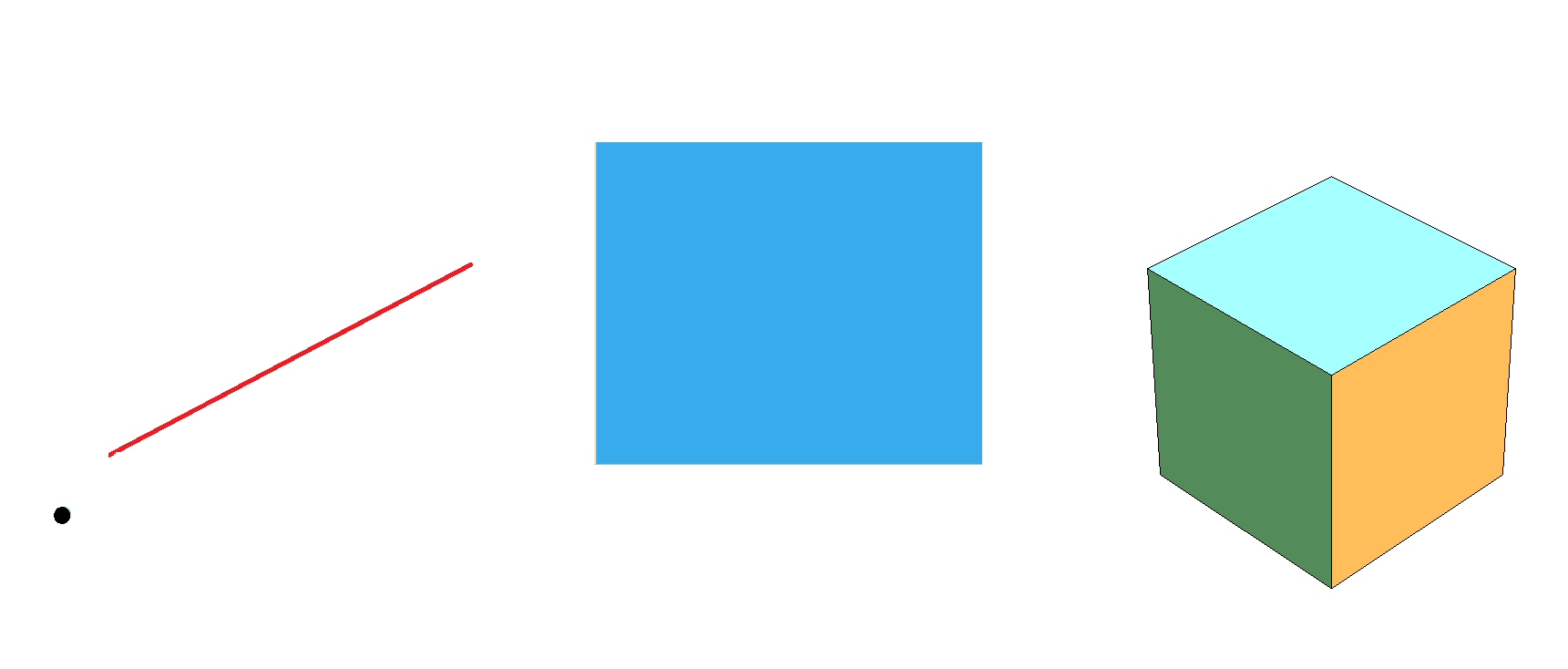}
\end{center}

\caption{A multivector $ M = a + \boldsymbol{v} + \iGA \boldsymbol{w} + \iGA t $ representing a point, line, area and volume, that can be added, subtracted, multiplied or divided by other multivectors. \label{MultiPic}}

\end{figure}

\subsubsection{Working with multivectors}

Addition and subtraction are simply defined by adding like components, that is, if $ M_1 = a_1 + \boldsymbol{v_1} + \iGA \boldsymbol{w_1} + \iGA t_1 $ and $ M_2 = a_2 + \boldsymbol{v_2} + \iGA \boldsymbol{w_2} + \iGA t_2 $, then  $ M_1 + M_2  = (a_1+a_2)+ (\boldsymbol{v_1}+\boldsymbol{v_2})+ \iGA (\boldsymbol{w_1}+\boldsymbol{w_2}) + \iGA (t_1+t_2) $ and similarly for subtraction.
The multiplication operation $ M_1 M_2 $ is simply formed by expanding the brackets containing each multivector, as for the product of two vectors. 

For the general multivector $ M = a + \boldsymbol{v} + \iGA \boldsymbol{w} + \iGA t $, it is useful to define two automorphisms.  Firstly reversion, that reverses the order of the basis products, giving $ {\reversion{M}} = a + \boldsymbol{v} - \iGA \boldsymbol{w} - \iGA t $ and space inversion $ M^{*} =  a - \boldsymbol{v} + \iGA \boldsymbol{w} - \iGA t $.
We can then define Clifford conjugation $ \cliffconj{M} = {\reversion{M}}^{*} = a - \boldsymbol{v} - \iGA \boldsymbol{w} + \iGA t $, that gives $ M \cliffconj{M} = a^2 - v^2 + w^2 - t^2 + 2 \iGA (a t - \boldsymbol{v} \cdot \boldsymbol{w} )$, a commuting complex-like number. We therefore find the inverse to $ M $ 
\be
M^{-1} = \cliffconj{M} /(M \cliffconj{M}).
\ee
The multivector inverse  fails to exist when $ M \cliffconj{M} = 0 $ or when $ a^2  + w^2 =  v^2 + t^2 $ and $ a t = \boldsymbol{v} \cdot \boldsymbol{w} $, which we can write as the single condition $ (\boldsymbol{v} + \iGA \boldsymbol{w})^2 = (a + \iGA t )^2 $.
The previously defined vector inverse in Eq.~(\ref{vectorInverse}) obviously now becomes a special case of this general multivector inverse.  This formula also applies as well in one and two dimensional space.

\subsubsection{The square root of a multivector}

In order to fully satisfy Descartes ideal of common algebraic operations being applicable to geometric quantities, such as lines and areas, we now also seek the square root of a multivector. 

Given a general two-dimensional multivector $ M = a + \boldsymbol{v} + \iGAT b $, then seeking a multivector $ N $, such that $ N^2 = M $, we find
\be \label{squareRoot}
N = M^{\frac{1}{2}} = \pm \frac{1}{2 c} \left ( 2 c^2 +  \boldsymbol{v} + \iGAT b \right ),
\ee
where we find from the quadratic formula that $ c^2 = \frac{a \pm \sqrt{a^2 - \boldsymbol{v}^2 + b^2 } }{2} $. 

However, if we use the {\it{amplitude}} of a multivector, defined as
\be
|M| = \sqrt{M \cliffconj{M}},
\ee
which is in general a complex-like number, then we can write the square root given in Eq.~(\ref{squareRoot}) as
\be \label{squareRoot2Dand3D}
M^{\frac{1}{2}} = \pm \frac{1}{\sqrt{M+\cliffconj{M} \pm 2 |M|}} \left ( M \pm |M| \right )
\ee
that allows a definition of the square root without the need to select individual components from the multivector under consideration.  It appears we may have defined the square root operation recursively, as the root operation also appears in the denominator.  However we reserve the square root symbol $ \sqrt{} $ for the case where the argument is always real or complex-like and so is already well defined using complex number theory.
Conveniently, we find that this result also applies unchanged in three dimensions, so that we therefore have a general result for the square root of a multivector up to and including dimension three~\cite{Chappell2015}.

Also, because multivector multiplication is associative we can now find all the rational  powers $ M^{p/2^q} $, where $ p,q $ are integers.   However, to take powers of multivectors, including fractional powers such as square roots, it is more general to achieve this through logarithms and exponents.

\subsection*{Exponential map of a multivector}

The exponential of a multivector is defined by constructing the Taylor series 
\be \label{exponentialMultivector}
\rme^M = 1 + M +\frac{M^2}{2!} + \frac{M^3}{3!} + \dots,
\ee
which is absolutely convergent for all multivectors $ M $ \citep{Hestenes111}.

Given a three-dimensional multivector $ a + \boldsymbol{v} + \iGA \boldsymbol{w} + \iGA t $, then defining $ F = \boldsymbol{v} +  \iGA \boldsymbol{w} $, we find $ F^2 = (\boldsymbol{v} + \iGA \boldsymbol{w})^2 = \boldsymbol{v}^2 - \boldsymbol{w}^2 + 2 \iGA \boldsymbol{v} \cdot \boldsymbol{w} $. We then define $  | F | = \sqrt{F \cliffconj{F}} = - \iGA \sqrt{F^2}  $ and so we can write $ F = \iGA \hat{F} | F | $, where $ \hat{F} = F/\sqrt{F^2} $ and $ \hat{F}^2 = 1 $.  Hence
\bea \label{derivationExpMultivector}
\rme^{a + \boldsymbol{v} + \iGA \boldsymbol{w} + \iGA t} & = & \rme^{a + \iGA t} \rme^{\iGA | F | \hat{F} } \\ \nonumber
& = & \rme^{a + \iGA t} \left ( 1 + j \hat{F} | F | - \frac{| F |^2 }{2!}- \frac{\iGA \hat{F} | F |^3}{3!} + \frac{| F |^4 }{4!} + \dots  \right ) \\ \nonumber
& = & \rme^{a + \iGA t} \left ( \cos | F | + \iGA \hat{F}  \sin | F |   \right ) . \nonumber
\eea
If $ | F | = 0 $, then referring to the second line of the derivation above, we see that all terms following $ \iGA \hat{F} | F | $ are zero, and so, in this case $ \rme^{a + \boldsymbol{v} + \iGA \boldsymbol{w} + \iGA t} = \rme^{a + \iGA t} (1+ \boldsymbol{v} + \iGA \boldsymbol{w} ) $. 

We can thus write a multivector in polar form 
\be
a + \boldsymbol{v} + \iGA \boldsymbol{w} + \iGA t  = r \rme^{\iGA \phi F/|F| } = r \left ( \cos \phi + \frac{F }{|F|} \sin \phi \right ),
\ee
where $ r = | M | $ and $ \phi = {\rm{arccosh}} \left ( \frac{a + \iGA t}{|M|} \right ) $. Hence
\be \label{LogMulti}
\log(a + \boldsymbol{v} + \iGA \boldsymbol{w} + \iGA t ) = \log |M| + \arccos \left ( \frac{a + \iGA t}{|M|} \right ) \frac{F }{|F|} .
\ee
This result being a generalization of the well known result for quaternions, when $ \boldsymbol{v} = t = 0 $.

We can now also define the multivector power
$ M^P = \rme^{ \log ( M ) P } $, where $ P $ can now also be generalized to a multivector, allowing many exotic possibilities.
For example, we could raise the multivector $ 2 + \hat{v} $ to the power of the unit vector $ \hat{v} $, giving
\be
(2+\hat{v})^{\hat{v}} = \rme^{ \log ( 2+\hat{v} ) \hat{v} } = 2 +\hat{v}  ,
\ee
using Eq.~(\ref{LogMulti}) and Eq.~(\ref{derivationExpMultivector}).

Using the exponential function, the full set of trigonometric and inverse trigonometric functions can now also be found for general multivector arguments\cite{Chappell2015}.  For example, we can take the cosine of a vector, giving $ \cos \boldsymbol{v} = \cosh | \boldsymbol{v} | $.

\subsection{The geometry of the multivector}

\begin{figure}[htb]

\begin{center}
\includegraphics[width=5.4in]{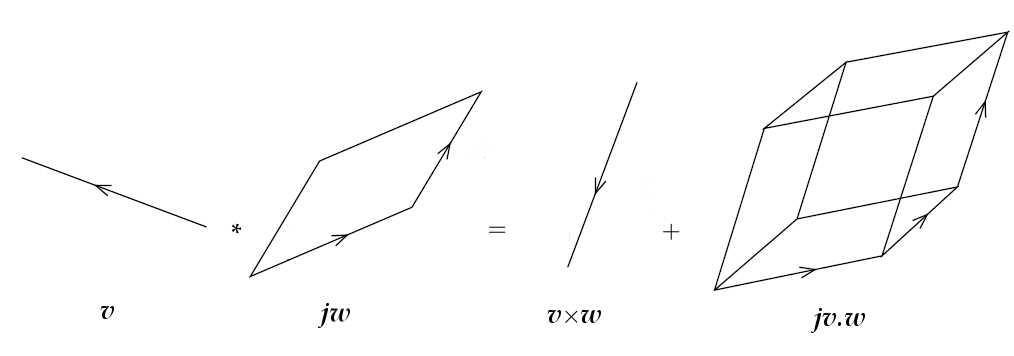}
\end{center}

\caption{A line $ \boldsymbol{v} $ multiplied by an area $ \iGA \boldsymbol{w} $. This produces a volume $ \iGA \boldsymbol{v} \cdot \boldsymbol{w} $ as expected but also a line $ - \boldsymbol{v} \times \boldsymbol{w} $ perpendicular to $ \boldsymbol{v} $ in the plane of $ \iGA \boldsymbol{w} $. \label{LineArea}}

\end{figure}

Having shown how the multivector, representing a set of geometric elements, consisting of a point, line, area and volume, is subject to the common algebraic operations, we can now ask some elementary geometrical questions such as: What is the result of multiplying a line by an area? We can calculate this for a line $ \boldsymbol{v} $ and a generally oriented area $ \iGA \boldsymbol{w} $ as $ \boldsymbol{v} \iGA \boldsymbol{w} = \iGA \boldsymbol{v} \wedge \boldsymbol{w} + \iGA \boldsymbol{v} \cdot \boldsymbol{w} = - \boldsymbol{v} \times \boldsymbol{w} + \iGA \boldsymbol{v} \cdot \boldsymbol{w} $.  As might have been expected this forms a volume $ \iGA \boldsymbol{v} \cdot \boldsymbol{w} $, and if the line is not perpendicular to the plane, we also produce a line, given by the vector $ - \boldsymbol{v} \times \boldsymbol{w} $ in the plane of $ \iGA \boldsymbol{w} $ and orthogonal to $ \boldsymbol{v} $, as shown in Fig.~\ref{LineArea}.

\subsection{Reflection of vectors}

Assuming a light ray with an incident vector $ \boldsymbol{a} $, is impinging on a plane mirror $ \iGA \hat{n} $, with a unit normal $ \hat{n} $, find the reflected vector.  We find the reflected vector, simply as follows,
\be
\boldsymbol{b} = - \hat{n} \boldsymbol{a} \hat{n} .
\ee

\begin{figure}[htb]

\begin{center}
\includegraphics[width=3.4in]{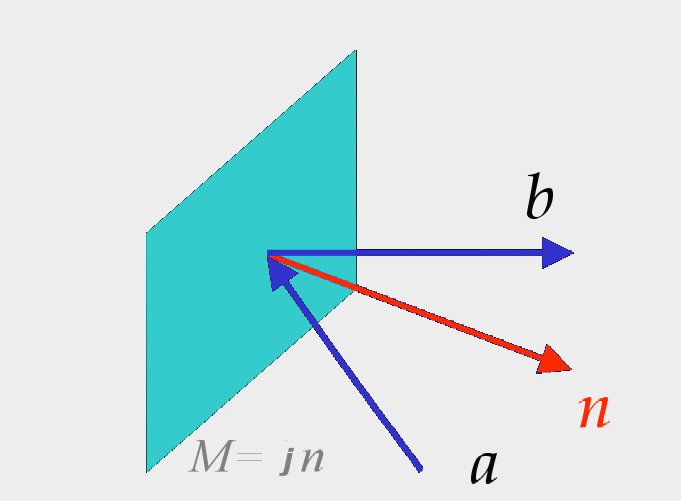}
\end{center}

\caption{A light ray incident on a plane mirror $ \iGA \hat{n} $.  We find the reflected ray $ \boldsymbol{b} = - \hat{n} \boldsymbol{a} \hat{n} $. \label{Mirror}}

\end{figure}

If we reflect $ \boldsymbol{b} $ in the same mirror we will recover the original vector $ \boldsymbol{a} $ because $  \hat{n}^2 = 1 $, however if we reflected $ \boldsymbol{b} $ in a slight rotated mirror plane $ \iGA \hat{m} $ we will in fact find a rotated vector
\be
\boldsymbol{b} =  \hat{m} \hat{n} \boldsymbol{a} \hat{n} \hat{m} = (  \hat{m} \cdot \hat{n} +  \hat{m} \wedge \hat{n} ) \boldsymbol{a} (  \hat{n} \cdot \hat{m} +  \hat{n} \wedge \hat{m} ) .
\ee
Now we have the unit bivector $ \hat{B} = \hat{m} \wedge \hat{n}/\sin \theta $ describing the plane of rotation and $ \theta =\arccos \hat{m} \cdot \hat{n} 
$ is the angle between the vectors $ \hat{m} $ and $ \hat{n} $, then
\be
\boldsymbol{b} =  ( \cos \theta + \hat{B}  \sin \theta ) \boldsymbol{a} ( \cos \theta - \hat{B}  \sin \theta ) = \rme^{\theta \hat{B} } \boldsymbol{a} \rme^{-\theta \hat{B} } ,
\ee
which will rotate the vector $ \boldsymbol{a} $ an angle of $ 2 \theta $ radians in the plane described by the unit bivector $ \hat{B} $.

\subsection{Rotation of vectors}

Therefore, if we wish to rotate a vector $ \boldsymbol{v} $ by an angle $ \theta $, then we can use the operation, 
\be \label{HamiltonRotations}
\boldsymbol{v}' = R \boldsymbol{v} \reversion{R} = \rme^{ \iGA \boldsymbol{w} \theta/2}  \boldsymbol{v} \rme^{ -\iGA \boldsymbol{w} \theta/2}
\ee
where $ R = \rme^{ \iGA \hat{w} \theta/2} $. The unit bivector $ \iGA \hat{w} $ sets the plane of rotation, with a perpendicular axis $ \hat{w} $, that rotates all vectors $ \theta $ radians within this plane.  In two dimensions this formula reduces to the single sided operator $ \boldsymbol{v}' = \rme^{ \iGAT \boldsymbol{w} \theta}  \boldsymbol{v} $ due to the anticommuting nature of $ \iGAT = e_{12} $ over vectors. The rotation formula in two dimension now analogous to the conventional formula for the rotation of vectors in the Argand plane.

If we allow a unit vector $ \boldsymbol{u} $ to represent the axis of a magnetic dipole.  
Then placing our particle in a magnetic field $ \boldsymbol{B} $, then using the rotation formula in Eq.~(\ref{HamiltonRotations}), we find the precession about the $ \boldsymbol{B} $ direction, given by
\be
\boldsymbol{u}' = \rme^{ \iGA \boldsymbol{B} t} \boldsymbol{u}  \rme^{ - \iGA \boldsymbol{B} t} ,
\ee
where we can see the precession is steady in time, and the rate of precession given by the strength of the field.

Rotations in geometric algebra are superior to orthogonal matrices in representing 3D rotations on four points: (i) it
is easier to determine the bivector representation of a rotation than the matrix representation, (ii) they avoid the problem of gimbal lock, (iii) it is more efficient to multiply bivectors than matrices, and (iv) if a
bivector product is not quite normalized due to rounding errors, then we simply divide by its norm, whereas if a product of orthogonal matrices is not orthogonal, then we need to use Gram-Schmidt orthonormalization, which is numerically expensive and not canonical.

\subsection{Interpreting solutions of quadratics using GA}

Imaginary numbers first appeared as the roots to quadratic equations, enabling solutions to equation such as $ x^2 + 4 = 0 $, with a solution $ x = 2 \sqrt{-1} $.  However, Gauss noted in 1825 that {\it{`The true metaphysics of the square root of minus one is elusive'}}.

However, with GA we can now supply a real geometrical solution to this equation, using the unit area $ \iGAT = e_{12} $, with $ x = 2 e_{12} $, that on substitution gives $ (2 e_{12})^2 + 4 = 0 $, that indeed solves the equation.  In fact many geometrical square roots of minus one exist, and in two dimensions we can write a general solution to $ M^2 = -1 $, as $ M = \boldsymbol{x} + \iGAT \sqrt{1 + \boldsymbol{x}^2 } $, where $ \boldsymbol{x} = x_1 e_1 + x_2 e_2 $.

Hence in GA, we can write a solution $ x = a + \iGAT b $, which from de~Moivre's theorem, gives $ R = r \rme^{\iGAT \theta } = r (\cos \theta + \iGAT \sin \theta ) $.
In two-space we can rotate vectors using the equation 
\be
 \boldsymbol{v}'= R \boldsymbol{v} ,
\ee
that will rotate a vector $ \boldsymbol{v} =v_1 e_1 + v_2 e_2  $ by $ \theta $ radians in a clockwise direction. For example, if $ R = \rme^{\iGAT \pi/2 } = \iGAT = e_{12} $, then if $ \boldsymbol{v} = e_2 $, then $ \boldsymbol{v}' = R \boldsymbol{v} = e_1 e_2 e_2 = e_1 $, or a clockwise rotation by $ \pi/2 $ degrees.  
Also, if we seek to rotate a vector by $ \pi $ radians then we find that $ \boldsymbol{v}'=  \rme^{ \iGAT \pi } \boldsymbol{v} = -  \boldsymbol{v} $.  This also illuminates the mysterious formula $ \rme^{\iGAT \pi } = -1 $, that simply means in this context, that rotating a vector by $ \pi $ radians flips its sign or in other words inverts its direction.

Hence solutions of quadratics using complex numbers imply that we are using rotation operators in the plane, instead of simply scaling along the real number line. This also explains why we always have two symmetrical complex solutions, if they exist, as they represent $ \pm \theta $ directions for the rotation operation.
So given a quadratic equation $ a x^2 + b x + c = 0 $, and substituting a rotor solution $ x = - r \rme^{- \iGAT \theta } $, we can, acting with the quadratic equation on a general vector $ \boldsymbol{v} $ on the left, produce the vector equation
\be
a r^2 \rme^{2 \iGAT \theta } \boldsymbol{v} - b r \rme^{ \iGAT \theta } \boldsymbol{v} + c \boldsymbol{v}  = 0 ,
\ee
where we used the property of exponentials that $ (\rme^{ \iGAT \theta})^2 = \rme^{2 \iGAT \theta} $.
Hence, in order to solve the quadratic these three vectors must sum to zero, that can be shown visually in Fig.~\ref{IsocellesQuadratic}, where, without loss of generality, we have chosen a reference direction $ \boldsymbol{v} = e_1 $.
\setlength{\unitlength}{1mm} 
\begin{figure}[tbp]
\begin{picture}(75,60)
\thicklines 
\put(55,52){\vector(-1,-1){45}} 
\put(10,7){\vector(1,0){67.2}}
\put(77.5,7){\vector(-1,2){22.4}}

\put(69,30){$ b r $} \put(44,2){$ c $}
\put(25,33){$ a r^2$} \put(67,11){$ \theta $} \put(52,42){$ \theta $} 
\end{picture}
\caption{Graphical solution to a quadratic equation. To solve graphically, we need to vary $ r \in \Re $ and $ \theta \in [ 0,\pi/2) $ while ensuring the arrows close in a triangle.  Note that we have an isosceles triangle and real solutions correspond to $ \theta = 0 $. } 
\label{IsocellesQuadratic}
\end{figure}
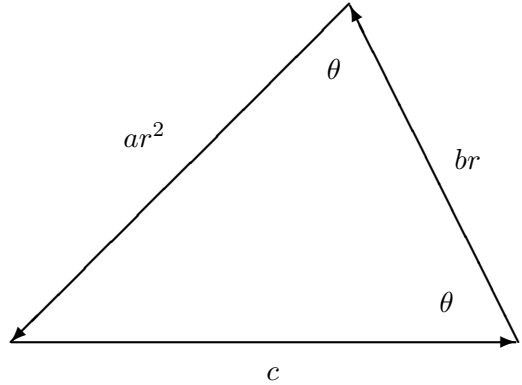
From Fig.~\ref{IsocellesQuadratic} we find $ r^2  = \frac{c}{a} $ that gives $ r = \sqrt{\frac{c}{a}} $ and $ \theta = \arccos \left ( \frac{b}{2 \sqrt{a c} } \right ) $.

\subsubsection{Quadratic equation example}

Assuming we are required to solve the quadratic
\be \label{quadraticExample}
x^2 + x+1 = 0
\ee
then we find the isosceles triangle shown in Fig. \ref{IsocellesQuadraticExample}.
\setlength{\unitlength}{1mm} 
\begin{figure}[tbp]
\begin{picture}(75,60)
\thicklines 
\put(55,52){\vector(-1,-1){45}} 
\put(10,7){\vector(1,0){67.2}}
\put(77.5,7){\vector(-1,2){22.4}}

\put(69,30){$ r $} \put(44,2){$ 1 $}
\put(25,33){$ r^2$} \put(67,11){$ \theta $} \put(52,42){$ \theta $} 
\end{picture}
\caption{Solving $ x^2 + x +1 = 0 $ graphically. From the isosceles triangle we see that $ r = \pm 1 $, and hence we have an equilateral triangle, which implies $ \theta = \pi/3 $.  Therefore $ x = - r \rme^{\pm \iGAT \theta} = -\frac{1}{2} \pm \iGAT \frac{\sqrt{3}}{2} $. }
\label{IsocellesQuadraticExample}
\end{figure}
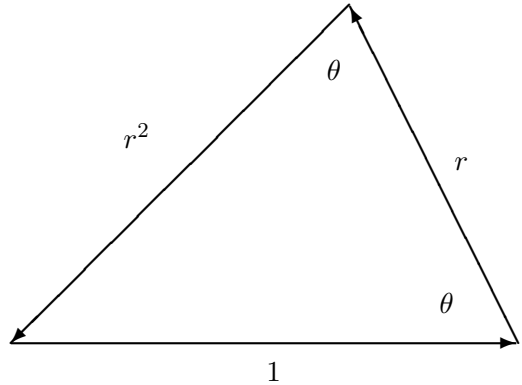
From the property of isosceles triangles we have $ r = 1 $ and we therefore realize, in this case, that we have an equilateral triangle, and hence $ \theta = \frac{\pi}{3} $, and hence we have the solution
\be
x = - \rme^{\pm \iGAT \pi/3 } = -\frac{1}{2} \pm e_{12} \frac{\sqrt{3}}{2}  \leftrightarrow -\frac{1}{2} \pm \sqrt{-1} \frac{\sqrt{3}}{2} ,
\ee
in agreement with the standard quadratic formula.  Using GA we therefore have a new method of solving quadratic equations.

Other extensions now present themselves for the quadratic equation, such as expanding the solution space further to allow $ x $ to be a quaternion (represented by bivectors), or to promote $ x $, and $ a,b,c $ to become full multivectors.  In the quaternion case, geometrically it represents the diagram in Fig. \ref{IsocellesQuadratic} being rotated out of the plane around the line $ c $.

\subsection{Calculus and elementary differentiation}

When applying calculus within the framework of GA, the one key addition is the need to respect the non-commuting nature of all products, however this allows many simplifications.
For example for the product rule between vectors (that now utilize the geometric product) we find
\be
(f g)' = f' g + f g'
\ee
the same as for algebraic variables, where the prime symbol represents differentiation with respect to some variable.  This now  replaces the two separate rules for the dot and cross products, that $ (f \cdot g)' = f' \cdot g + f \cdot g' $ and $ (f \times g)' = f' \times g + f \times g' $.
For the more general case of the vector gradient, defined by $ \boldsymbol{\nabla} = \ex \frac{\partial}{\partial x} + \ey \frac{\partial}{\partial y} + \ez \frac{\partial}{\partial z} $, we find for a vector field $ \boldsymbol{v} = v_1 e_1 + v_2 e_2 + v_3 e_3 $ that
\be
\boldsymbol{\nabla} \boldsymbol{v} = \boldsymbol{\nabla} \cdot \boldsymbol{v} + \iGA \boldsymbol{\nabla} \times \boldsymbol{v}
\ee
forming a convenient union of the divergence and curl.  However there is no reason why we cannot apply the gradient to a general multivector field, as $ \boldsymbol{\nabla} M $, where $ M $ is a three-space multivector.

\subsection{Maxwell's equations in GA}

Electromagnetism is one of the foundational theories of physics and Maxwell's equations \cite{Maxwell:1865} were first published in 1865.  Maxwell's original equations were written for three-space, requiring 12 equations in 12 unknowns. These equations were later rewritten by Heaviside and Gibbs, in the then newly developed formalism of dot and cross products, which reduced them to the four equations now seen in most modern textbooks  \cite{Griffiths:1999} and shown below in S.I. units
\bea \label{MaxwellClassical}
\boldsymbol{\nabla} \cdot \boldsymbol{E} & = & \frac{\rho}{\epsilon},   \,\,\,\,\,\,\, \rm{(Gauss\text{'}s \,\, law)} ; \\ \nonumber
\boldsymbol{\nabla} \times \boldsymbol{E} +  \partial_t \boldsymbol{B}  & = &  0 , \,\,\,\,\,\,\,\,\, \rm{(Faraday\text{'}s \,\, law)} ; \\ \nonumber
\boldsymbol{\nabla} \times \boldsymbol{B} - \frac{1}{c^2} \partial_t \boldsymbol{E} & = & \mu \boldsymbol{J},  \,\,\,  \rm{(Amp\grave{e}re\text{'}s \,\, law)} ; \\ \nonumber
\boldsymbol{\nabla} \cdot \boldsymbol{B}  & = &  0 , \,\,\,  \,\,\,  \,\,\,  \rm{(Gauss\text{'}s \,\, law \,\, of \,\, magnetism)} ,   \nonumber
\eea
where $ \boldsymbol{E}, \boldsymbol{B}, \boldsymbol{J} $ are conventional three-vector fields, with $ \boldsymbol{E} $  the electric field strength and $ \boldsymbol{B} $ the magnetic field strength and $ \boldsymbol{\nabla} $ is the three gradient defined previously.

However inspecting the form of the geometric product for $ \boldsymbol{u} \boldsymbol{v} = \boldsymbol{u} \cdot \boldsymbol{v}  + \iGA \boldsymbol{u} \times \boldsymbol{v} $, we can see that these equations can now be combined. If we multiply the second and fourth equations by  $ \iGA $, then the first and second equations can be combined along with the third and fourth to give
\bea \label{MaxwellClassicalWithiTwoEqn}
\boldsymbol{\nabla} \boldsymbol{E} +  \partial_t \iGA \boldsymbol{B} & = & \frac{\rho}{\epsilon}  \\ \nonumber
\boldsymbol{\nabla} \iGA \boldsymbol{B} +  \frac{1}{c^2} \partial_t \boldsymbol{E} & = & -\mu \boldsymbol{J},  \nonumber
\eea
where $ \boldsymbol{\nabla} \boldsymbol{E} = \boldsymbol{\nabla} \cdot \boldsymbol{E} + \iGA \boldsymbol{\nabla} \times \boldsymbol{E} $.
However, these two remaining equations can now be added to produce
\be \label{MaxwellClassicalWithiOneEqn}
\left (\frac{1}{c} \partial_t + \boldsymbol{\nabla} \right ) (  \boldsymbol{E} +  \iGA c \boldsymbol{B} ) =  \frac{\rho}{\epsilon} -c \mu \boldsymbol{J}.
\ee
If we define the electromagnetic field $ F = \boldsymbol{E} +  \iGA c \boldsymbol{B} $ and the four-gradient $ \partial = \frac{1}{c} \partial_t + \boldsymbol{\nabla} $, with the source $ J =  \frac{\rho}{\epsilon} -c \mu \boldsymbol{J} $, we find
\be \label{boxMaxwell}
\partial F = J.
\ee
We can see that the $ \boldsymbol{B} $ field, is written as a pseudovector $ \iGA \boldsymbol{B} $, as part of the field $ F $.  The different nature of the $ \boldsymbol{E} $ and $ \boldsymbol{B} $ fields is thus evident from the GA formalism but obscured in the tensor and Gibbs' vector formalism, where both are represented as simply polar vectors.

Also if we wish to describe Maxwell's original four equations as shown in Eq.~(\ref{MaxwellClassical}) in plain English we would find it very cumbersome, however with GA, inspecting Eq.~(\ref{boxMaxwell}), we can simply say that the gradient of the field $ F$ observed is proportional to the electromagnetic sources present.

\subsection{The Dirac equation}

The Dirac equation is the relativistic wave equation describing spin-$\frac{1}{2}$ particles.
We find using GA, that we can write the free Dirac equation in real three-space as
\be \label{freeDirac}
\partial F = - m F^{*} \iGAT ,
\ee
where the field $ F $ is now the full multivector $ F = a +\boldsymbol{E} + \iGA \boldsymbol{B} + \iGA b $, and $ F^{*} $ signifies space inversion of the multivector $ F $ and $ \iGAT =  e_1 e_2 $  defined earlier.

The similarity of Dirac's equation with Maxwell's equation now becomes evident, comparing Eq.~(\ref{freeDirac}) and Eq.~(\ref{boxMaxwell}).  Also setting $ m = 0 $ in the Dirac equation, in accordance with a massless photon,  we find $ \partial F = 0 $, which is the source free Maxwell equation.
Also, with the GA form of the Dirac equation, we can see that it describes a multivector field, given by Eq.~(\ref{generalMultivector}), over real three-space, that is, at each point in three-space we have a multivector valued field defined.
This is clearly a significantly simplified representation of the Dirac equation, which is normally considered embedded in four-dimensional spacetime employing  $4 \times 4 $ complex matrices, as shown in Appendix B.

\subsection{Spacetime}

It might be argued that the correct framework for physical theories is Minkowski spacetime that includes a fourth dimension for time.  However, this can be once again absorbed into GA in three dimensions without the addition of an extra dimension.  For simplicity we illustrate here the case for two dimensions, though, because most interactions typically considered in special relativity are planar, it is still practically very useful.  If we define a spacetime event multivector as
\be
S = \boldsymbol{x} + i t
\ee
where $ \boldsymbol{x} = x_1 e_1 + x_2 e_2 $ is the space coordinate and $ t $ is the time, where $ i = e_{12} $, the bivector of the plane.
We then find that 
\be
S^2 = (\boldsymbol{x} + i t)(\boldsymbol{x} + i t) = \boldsymbol{x}^2 -t^2 + t (\boldsymbol{x} i + i \boldsymbol{x} ) = \boldsymbol{x}^2 -t^2 
\ee
using the fact that $ i \boldsymbol{x} $ is anticommuting, thus producing the conventional spacetime distance, with the negative contribution to the metric from the time component.
If we define the Lorentz transformation as $ L = \rme^{\phi i \hat{\boldsymbol{v}} } \rme^{i \theta } $, applied to the spacetime coordinates using the transformation $ S' = L S \cliffconj{L} $, where $ \cliffconj{L} = \rme^{-i \theta } \rme^{-\phi i \hat{\boldsymbol{v}} }  $, we find
\be \label{coordinateBoosts}
(S')^2 = \rme^{\phi i \hat{\boldsymbol{v}} } \rme^{i \theta } S \rme^{-i \theta } \rme^{-\phi i \hat{\boldsymbol{v}} }  \rme^{\phi i \hat{\boldsymbol{v}} } \rme^{i \theta } S \rme^{-i \theta } \rme^{-\phi i \hat{\boldsymbol{v}} }  = \rme^{\phi i \hat{\boldsymbol{v}} } \rme^{i \theta } S^2 \rme^{-i \theta } \rme^{-\phi i \hat{\boldsymbol{v}} } = S^2,
\ee
using the fact that $ S^2 $ is a scalar, thus leaving the spacetime distance invariant, defining the restricted Lorentz group \citep{zeni1992thoughtful}. When applied to a spacetime event $ S $, we find for $ \phi = 0 $ pure rotations, and for $ \theta = 0 $, we find pure boosts.  This example used to briefly illustrate the capacity of GA to embody the properties of spacetime without the requirement of an additional dimension \cite{chappell2011revisiting}.

\section{Conclusion}

In this paper we ask the question `What is the simplest and most natural mathematical representation of three-dimensional physical space?' and we illustrate that GA provides a natural formalism subsuming the algebra of complex numbers and quaternions into a single algebraic system over a real field.  We provide pedagogical examples illustrating that  Clifford's GA is an elegant mathematical system that successfully represents the key properties of physical space, such as points, lines, areas and volumes as well as a simplified representation of physical theories with the two examples of Maxwell's equations and the Dirac equation, both being written as single equations in real three dimensional space.  
We also show how the multivector, shown in Eq.~(\ref{generalMultivector}), can be viewed as a generalized number, useful in representing different physical and geometrical quantities, such as electromagnetic fields, but still subject to the basic algebraic operations of addition, subtraction, multiplication, division and square root.   

At a more elementary level, we show how a general quadratic equation can be solved without recourse to complex numbers, giving the solutions geometric meaning as rotations in the plane.

We adopt the symbols $ \iGAT = e_{12} $ in two dimensions and $ \iGA = e_{123} $ for three dimensions as two geometric replacements for the generic scalar unit imaginary $ \sqrt{-1} $, and we note a distinction between $ \iGAT $ and $ \iGA $ in the GA form of the Dirac equation in Eq.~(\ref{freeDirac}).  

The development of GA is now expanding rapidly, with benefits being found in research into quantum field theory \cite{Dolby2001}, quantum tunneling \cite{Doran1997}, quantum computing \cite{DoranParker:2001}, spacetime \cite{Hestenes2003,rowland2010,GA2}, general relativity and cosmology \cite{berrondo2012,Doran1993,Doran2005}, computer vision \cite{Doran1998}, protein folding \cite{Chys2012}, optics and metamaterials \cite{Santos2012,Sugon2004,Matos2010}, conformal algebra \cite{HuHao2012}, electrodynamics \cite{vold1993}, electrical circuit analysis \cite{Castro2012} and EPR-Bell experiments \cite{Christian2012}.

Many commentators believe that Clifford's mathematical system {\it{`should have gone on to dominate mathematical physics'}} \cite{Doran2003}, but, Clifford died young, at the age of just 33 and vector calculus was heavily promoted by Gibbs and rapidly became popular, eclipsing Clifford's work, which in comparison appeared strange with its non-commuting variables.  With the benefit of hindsight, the non-commuting nature of GA reflects the non-commutivity of rotations in three-space, and hence is exactly what is required for these variables.  
Gibbs' system of vectors was relatively efficient with regard to Maxwell's equations, but with the new scientific discoveries of quantum mechanics and relativity it was found that standard vector analysis needed to be supplemented by many other mathematical techniques such as: tensors, spinors, matrix algebra, Hilbert spaces, differential forms etc.~and as noted in \cite{Simons:2009}, {\it{`The result is a bewildering plethora of mathematical techniques which require much learning and teaching, which tend to fragment the subject and which embody wasteful overlaps and requirements of translation.'}}  Conversely as we have seen GA is a natural formalism for not only Maxwell's equations, Eq.~(\ref{boxMaxwell} and also quantum mechanics, Eq.~(\ref{freeDirac}, but also special relativity \cite{chappell2011revisiting}.

With regard to an educational setting, we also have shown that geometric algebra provides a natural representation of the basic properties of physical space, allowing intuitive manipulation of lines, areas and volumes using elementary algebraic operations, such as addition and multiplication. Vectors can now be treated like normal algebraic quantities that also have an inverse, with the added simplification that the dot and cross products do not need to be separately defined but are produced as a byproduct from the geometric product.  Hence it appears to be an excellent formalism to introduce into high-school and undergraduate university curricula as a powerful tool for basic geometrical analysis of space, which can be naturally extended to the study of university level subjects in electromagnetism, quantum theory and special relativity.

\section{appendix}

\subsection{Dirac equation}

Dirac extended Schr$\ddot{\rm{o}}$dinger's and Pauli's equation into a relativistic setting in 1928, producing the equation
\be
\gamma^{\mu} \partial_{\mu} \psi = -m  \psi {\rm{i}} ,
\ee
where $ {\rm{i}} = \sqrt{-1} $ and which uses the Einstein summation convention, where
\be
\gamma^{0} = \begin{bmatrix} 1 & 0 & 0 & 0 \\
0 & 1 & 0 & 0 \\
0 & 0 & -1 & 0 \\
0 & 0 & 0 & -1  \end{bmatrix}  , \gamma^{1} = \begin{bmatrix} 0 & 0 & 0 & 1 \\
0 & 0 & 1 & 0 \\
0 & -1 & 0 & 0 \\
-1 & 0 & 0 & 0  \end{bmatrix} , \gamma^{2} = \begin{bmatrix} 0 & 0 & 0 & -{\rm{i}} \\
0 & 0 & {\rm{i}} & 0 \\
0 & {\rm{i}} & 0 & 0 \\
-{\rm{i}} & 0 & 0 & 0  \end{bmatrix}  , \gamma^{3} = \begin{bmatrix} 0 & 0 & 1 & 0 \\
0 & 0 & 0 & -1 \\
1 & 0 & 0 & 0 \\
0 & -1 & 0 & 0  \end{bmatrix} .
\ee
The gamma matrices satisfy the relation
and
\be
\{ \gamma^{\mu} , \gamma^{\nu} \} = 2 g^{\mu \nu} ,
\ee
as expected for a set of orthonormal basis vectors.
Hence, the opinion of many people, that Dirac rediscovered Clifford's geometric algebra with its anti-commuting basis vectors.
Dirac's complex, four-space equation using $ 4 \times 4 $ complex matrices, is isomorphic to the real three-space version, shown in Eq.~(\ref{freeDirac}).

\bibliographystyle{model1a-num-names}

\bibliography{quantum}

\end{document}